\documentclass{pasa}
\usepackage[authoryear]{natbib}
\bibpunct{(}{)}{;}{a}{}{,}
\usepackage{aas_macros}
\usepackage{hyperref} 
\hypersetup{colorlinks,citecolor=blue,linkcolor=blue,urlcolor=blue}

\def\aegean{{\sc Aegean}}
\def\bane{{\sc BANE}}
\def\mimas{{\sc MIMAS}}
\def\aeres{{\sc AeRes}}

\def\deg{\ensuremath{^\circ}}

\title[Aegean v2.0]{Source finding in the era of the SKA (precursors): \aegean{} 2.0}
\author[Hancock et al.]{Paul J. Hancock$^{1,2}$\thanks{email: Paul.Hancock@curtin.edu.au}, Cathryn M. Trott$^{1,2}$, Natasha Hurley-Walker$^{1}$\\
\affil{$^1$International Centre for Radio Astronomy Research, Curtin University, Bentley, WA 6102, Australia}%
\affil{$^2$ARC Centre of Excellence for All-sky Astrophysics (CAASTRO)}}%

\jid{PASA}
\doi{10.1017/pas.\the\year.xxx}
\jyear{\the\year}

\begin{document}

\begin{abstract}
In the era of the SKA precursors, telescopes are producing deeper, larger images of the sky on increasingly small time-scales.
The greater size and volume of images place an increased demand on the software that we use to create catalogues, and so our source finding algorithms need to evolve accordingly.
In this paper we discuss some of the logistical and technical challenges that result from the increased size and volume of images that are to be analysed, and demonstrate how the \aegean{} source finding package has evolved to address these challenges.
In particular we address the issues of source finding on spatially correlated data, and on images in which the background, noise, and point spread function, vary across the sky.
We also introduce the concept of forced or priorized fitting.
\end{abstract}
\begin{keywords}
radio astronomy -- software -- source-finding -- keyword(N)
\end{keywords}
\maketitle%

\section{Introduction}
\label{sec:intro}

The primary goal of any source finding program is the accuracy of the resulting catalogues.
However, sensitive, wide-field telescopes with fast imaging capabilities, mean that additional requirements are being placed on the software that supports these instruments.
Thus source finding programs need to evolve with the data which they are processing.
Source finding programs are increasingly being required to: make effective use of hardware in order to run in near-real time, integrate with a range of software ecosystems, and have the flexibility to support a range of science goals.
We group these demands into three categories: correctness, hardware utilization, and interoperability. 

Time domain astronomy places demands on each of these categories and is thus a good exemplar case.
Time domain astronomy is an area of great interest to many of the new radio telescopes either in operation (MWA \citet{Tingay_murchison_2013}, LOFAR \citet{Rottgering_lofar_2003}), in commissioning (ASKAP \citet{Johnston_science_2008}), under construction (MeerKAT \citet{Jonas_meerkat_2009}), or in planning (SKA\footnote{\href{http://www.skatelescope.org}{www.skatelescope.org}}).
Studies of variable and transient events are almost entirely focused on variations of source flux density with time.
For example the VAST project \citep{Murphy_vast_2013,banyer_vast:_2012} and the the LOFAR Transients key science project \citep{Swinbank_lofar_2015} have both created analysis pipelines to detect variable and transient objects from light curve data.
The critical question for such work is then: given a sequence of flux densities and associated $1\sigma$ error estimates, what is the degree of variability exhibited by each source, and how confident are we that each source is varying or not.
The process of creating light-curves from catalogue data involves cross-matching sources between different epochs.
Accurate cross-matching requires that sources positions and uncertainties are correctly measured and reported.
Since false positives in the detection process appear as single epoch transients, the reliability of a source finding program is of great importance.
Determining the degree and confidence of variability depends critically on the uncertainty associated with each flux measurement in a light curve, and thus the accurate reporting of uncertainties is a priority.
In short, detecting variable and transient sources requires that the underlying source-finding process can be relied upon.
This in turn requires that the image background and noise, the synthesized beam, and the degree to which the data are correlated, are all important in the source-finding and characterization process.
Time domain astronomy is thus limited by the correctness of the catalogues which are extracted from images.

In order to trigger useful follow up observations of transients, the time between observation and detection needs to be minimised.
This means that the speed of a source finding program can be critical, and so the ability to utilize multiple CPU or GPU cores becomes important.

Finally a source finding program will always comprise only a single component in a larger processing pipeline, and so the ease with which the program can be incorporated into this environment can be the difference between a practical and impractical solution.

\aegean{}\,2.0 aims to address the issues of correctness, hardware utilization, and interoperability.
Hardware utilization for \aegean{} has been improved by using multiple processes for the fitting stage.
Interoperability has been improved by exposing the core \aegean{} functionality in a library called AegeanTools, and by allowing for a greater variety of input and output catalogue formats.
These two topics are important but their development is not novel and so we do not discuss them in detail.
In this paper we focus on issues of correctness of the output catalogue.

\section{Why 2.0?}
\label{sec:why}

\aegean{} \citep{Hancock_compact_2012} was developed to tackle some of the shortcomings of the software that was in common use among radio astronomers in c.\,2010.
\aegean{} has been upgraded and improved upon since 2012, thanks to input from work such as \citet{Huynh_completeness_2012} and \citet{Hopkins_askapemu_2015}, however these works focus on simulated images of modest sizes.
Running \aegean{} on data from SKA pathfinders such as MWA/LOFAR/ASKAP, embedded within scientific work-flows, a number of issues have come to light that have needed to be addressed.
These issues are related to practical and theoretical requirements inherent in wide-field radio images, and are invisible to the end user of a radio source catalogue.
The requirements include the ability to:
\begin{enumerate}
\item correctly fit a model to data which are spatially correlated,
\item work on large fields of view where many parameters vary across the image,
\item spread processing across multiple cores/nodes in a high performance computing (HPC) environment, and
\item integrate into a variety of work-flows.
\end{enumerate}

Whilst some source finding packages address a subset of these issues, prior to \aegean{}\,2.0 no one package was able to address all four at the same time.
This paper serves as a point of reference for the many developments of \aegean{} as well as a more formal description of the new capabilities.

During the production of the GaLactic and Extragalactic All-sky Murchison Widefield Array (GLEAM) survey catalogue \citep{Hurley-Walker_galactic_2017}, all four of the above issues came to bear, and thus \aegean{} was updated accordingly.

In the sections that follow we focus on: the effects of correlated data, estimating bias and uncertainty (\S\,\ref{sec:nlls}),  estimation of background and noise properties (\S\,\ref{sec:bane}), incorporating a variable point spread function (\S\,\ref{sec:psf}), the process of priorized fitting (\S\,\ref{sec:priorized}), extended source models (\S\,\ref{sec:island}), and sub-image searching (\S\,\ref{sec:regions}).
We conclude with a summary (\S\,\ref{sec:summary}) and future development plans (\S\,\ref{sec:future}).

The three programs discussed in this paper (\aegean, \bane, and \mimas{}) are part of the AegeanTools software suite.
AegeanTools is available for download from \href{https://github.com/PaulHancock/Aegean}{GitHub}, along with a user guide and application programming interface (API).

\section{Test data}
\label{sec:data}
Throughout this work we rely on two test data sets: simulated data, and observational data.

A simulated test image was created with the following properties:
\begin{itemize}
\item The image centre is at $(\alpha=180\deg,\delta=-45\deg)$ (the assumed zenith),
\item The image size is $10k\times 10k$ pixels, comparable to the mosaics generated for the GLEAM survey,
\item Pixel area is $0.738'\times0.738'$ at the image centre,
\item Projection is zenithal equal area (ZEA) in order to accurately represent the large area of sky covered,
\item Pixels below $\delta=-84\deg$ are masked (blank),
\item The point spread function of the image changes with sky coordinates, being circular at the image centre and elongating with increasing zenith angle,
\item The image noise varies from 0.1 to 0.2 Jy/beam, 
\item A large scale smooth background varies from -0.5 to 0.5 Jy/beam,
\item A population of point sources were injected with a peak flux distribution ranging from 0.1 to 1000\,Jy, a source count distribution of $N(S)\propto S^{3/2}$, and sky density of 14.5\,$\textrm{deg}^{-2}$, and
\item The source population is uniformly distributed in $(\alpha,\delta)$.
\end{itemize}
The simulated image can be downloaded from Zenodo \citep{hancock_image_2017}, and the code to generate the test image and all figures used in this paper can be found on the \href{https://github.com/PaulHancock/AegeanPaper2.0Plots}{AegeanPaper2.0Plots} GitHub repository.

\begin{figure}
\includegraphics[width=0.9\linewidth]{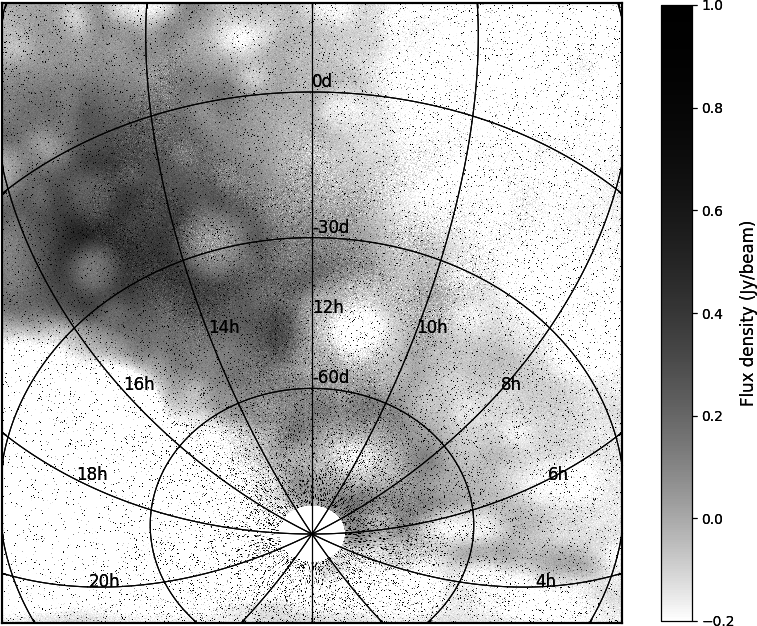}
\includegraphics[width=0.9\linewidth]{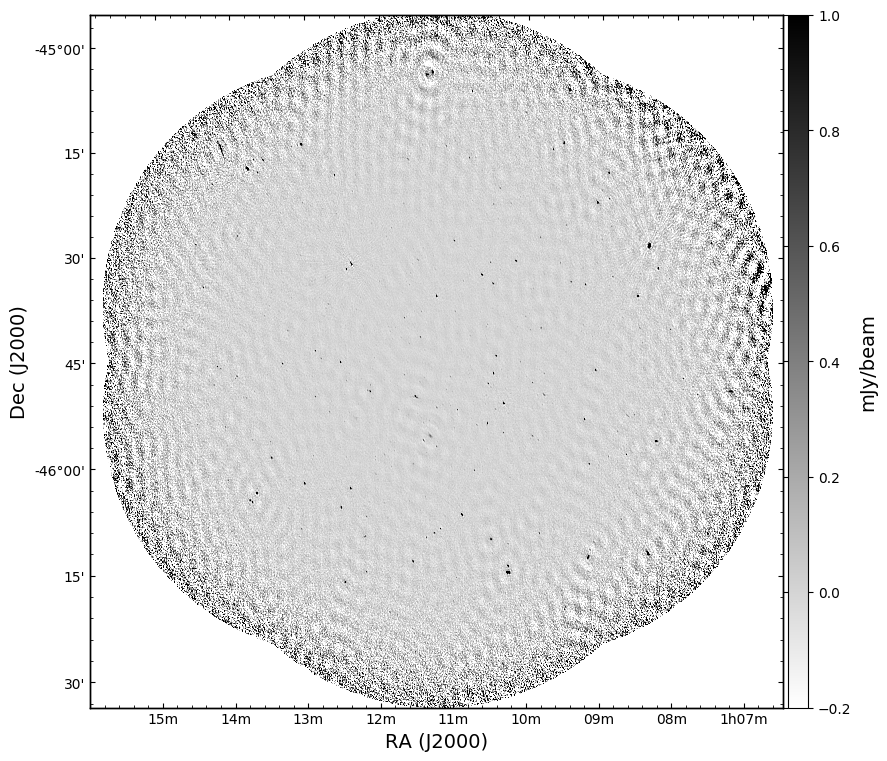}
\caption{Test data used in this paper.
{\em Upper} The simulated test image. The color scale has been chosen to exaggerate the large scale background emission. The injected sources appear as black points.
{\em Lower} An image from the Phoenix deep field, epoch of 1997. 
}
\label{fig:data}
\end{figure}

The observational data are taken from \citet{Hancock_radio_2016,Hancock_data_2016}.
We use just the 1997 epoch of observations originally observed as part of the Phoenix project \citep{hopkins_phoenix_1998}.
These data represent real observations, and include calibration errors, un-cleaned sidelobes, and a background and noise that changes throughout the image.
This data set will allow us to test the performance of \aegean{} in good but non-ideal conditions.
The simulated and observational images are shown in Figure\,\ref{fig:data}.

\section{Least-squares fitting}
\label{sec:nlls}
The aim of least squares minimisation is to minimise the following target function:
\begin{equation}
T = \left(\mathrm{model}-\mathrm{data}\right)^T\left(\mathrm{model}-\mathrm{data}\right)
\label{eq:target}
\end{equation}

Ordinary least-squares is the best linear unbiased estimator for data in which the errors have zero expectation and the data are uncorrelated.
Radio images have spatially correlated pixels and thus ordinary least-squares is no longer unbiased.
The source models that are typically fit (elliptical Gaussians) are non-linear models, and thus iterative methods are required to perform the fit.
Even with uncorrelated data, fitting non-linear models results in parameters being biased and correlated.

Knowing the degree to which the data are correlated we can alter the target function in equation 1, to be:
\begin{equation}
 T^\prime = \left(\mathrm{model}-\mathrm{data}\right)^TC^{-1}\left(\mathrm{model}-\mathrm{data}\right)
 \label{eq:tprime}
\end{equation}
where the matrix $C$ is the covariance matrix.
This expression is unbiased for data distributed as a generalized multivariate Gaussian.
Equation \ref{eq:target} is correct when $C=\sigma^2I$, where $I$ is the identity matrix, and $\sigma^2$ is the variance of the data.
This modification of the target function will remove the bias introduced by having correlated data, but will not affect the bias due to non-linear models.
We will first discuss how a covariance matrix can be incorporated into current minimisation libraries (\ref{sec:nlls:cinv}), and then discuss ways to estimate the uncertainty and bias in the resulting parameters (\ref{sec:nlls:errors}).

\subsection{Including the covariance matrix}
\label{sec:nlls:cinv}
One of most commonly used fitting libraries is MINPACK \citep{More_MINPACK_1984}. 
The MINPACK program LMFIT uses the Lavenburg-Marquardt algorithm to perform non-linear least squares fitting, with the target function as per Eq\,\ref{eq:target}. 
There is no functionality within the MINPACK library to include a modified version of the target function as per Eq\,\ref{eq:tprime}. 
The LMFIT function requires a pointer to a user-generated function which must return the vector $\left(\mathrm{model}-\mathrm{data}\right)$, and it is the Euclidean norm of this vector that is then minimised. 

The problem that we are faced with is constructing a vector $X$ such that $||X|| = X^TX = T^\prime$. With this modified vector it is then possible to use pre-compiled, debugged, optimized, and well tested code to fit our data.

\subsubsection*{Calculating the covariance matrix}

The covariance of the data in a radio image is due to one of two effects depending on the type of instrument used.
A single dish map will have pixels that are correlated due to the primary beam of the instrument, whilst a map made from an interferometer will have pixels that are correlated due to the synthesized beam.
In either case the covariance matrix can be easily calculated using knowledge of the primary and/or synthesized beam.
A typical single dish telescope has a primary beam that is well approximated locally by a Gaussian.
An interferometer will have a synthesized beam that is some combination of sinc functions, depending on the $(u,v)$ sampling and weighting functions.
Ideally the inverse covariance matrix would match the synthesized or primary beam, projected onto the pixel coordinates in the image.
In many cases the synthesized beam can be approximated locally by an elliptical Gaussian.
For a Gaussian beam with shape parameters of ($\sigma_x,\sigma_y,\theta$), we have a covariance matrix which is:

\begin{equation}
\begin{split}
C_{i,j} = & \exp\left[-\left(\delta x\sin\theta + \delta y\cos\theta\right)^2/2\sigma_x^2\right] \\ 
\times &\exp\left[-\left(\delta x\cos\theta - \delta y\sin\theta\right)^2/2\sigma_y^2 \right]\\
\delta x = & x_i - x_j \\
\delta y = & y_i - y_j
\end{split}
\end{equation}

The covariance matrix is thus real valued and symmetric.
The inverse of this matrix exists for all values of ($\sigma_x,\sigma_y,\theta$), however for large ($\sigma_x,\sigma_y$) the inverse matrix is numerically unstable.
In radio astronomy applications common practice is to use a restoring beam that has a full width half maximum of between 3-5 pixels, corresponding to a $\sigma$ of between $\sim 1.5-2.2$. 
For these values of $\sigma$ the covariance matrix is easily inverted.

In order to incorporate the inverse covariance matrix into the LMFIT routine, we seek a matrix $B$ such that $(BX)^TBX = X^TC^{-1}X$.
This is true if $B^TB = C^{-1}$.

\subsubsection*{Calculating the matrix B}
\label{sec:nlls:bmat}
The required matrix $B$ is a root of the matrix $C^{-1}$, however many such roots exist.
For computational reasons we also require that $B$ is real valued.
Since $C^{-1}$ is real valued, then one of its roots must also be real valued.
Using eigen-decomposition we can construct a matrix $Q$ of the eigen-vectors of $C$ and a diagonal matrix $\Lambda$ of the eigen-values of $C$, such that 
\begin{equation}
(Q\Lambda)^{-1}Q\Lambda = C
\label{eq:qlambda}
\end{equation}
The eigen-values of $C$ are all positive so that we can create a new matrix $\Sigma$ such that 
\begin{equation}
\Sigma_{i,i} = \frac{1}{\sqrt{\Lambda_{i,i}}}
\label{eq:sigma}
\end{equation}
This then gives the identity $(Q\Sigma)^{-1}Q\Sigma = C^{-1}$, which means that $B=Q\Sigma$ is a positive square root of the matrix $C^{-1}$ as required.
The matrix $B$ is unitary and real so that $B^T=B^{-1}$.
This choice of $B$ then gives $$(BX)^TBX = X^TB^TBX = X^T(Q\Sigma)^{-1}Q\Sigma X = X^TC^{-1}X$$ as required.

In practice, the matrix $C$ may have small negative eigen-values, when the synthesized beam is large compared to the pixel size, or when the number of pixels being fit is large.
In either case the matrix $\Sigma$ is modified to be:
\begin{equation}
\Sigma_{i,i} = \frac{1}{\sqrt{|\Lambda_{i,i}|}}
\label{eq:sigmamod}
\end{equation}

\subsection{Parameter uncertainty and bias}
The minimisation functions provided by MINPACK will return the parameter estimates that minimise the sum of the squares of the vector $X$, as well as a variance matrix.
If we provide the matrix $BX$ as described in the previous sections, then the parameter estimates will not be biased by the correlation in the input data.
However the returned variance (or $1\sigma$ error) estimates will not incorporate the effects of the correlated data.
Here we discuss the additional uncertainty and bias that is caused by having a non-linear model and correlated data.

\subsubsection*{Uncertainty}
\label{sec:nlls:errors}
The variance of the resulting fitted parameters is related to the Fisher Information Matrix by: $\sigma_i^2 = \left(F^{-1}\right)_{i,i}$.
The Fisher Information Matrix for a real-valued generalized Gaussian is given by\citep{vantrees_detection_1947}:
\begin{equation}
F_{i,j} = \frac{\partial{G}}{\partial x_i}C^{-1}\frac{\partial{G}}{\partial x_j} 
+ \frac{1}{2}\mathrm{tr}\left(C^{-1}\frac{\partial{C}}{\partial x_i}C^{-1}\frac{\partial{C}}{\partial x_j}\right)
\label{eq:fimfull}
\end{equation}
Where $G$ is the model of interest (an elliptical Gaussian), and $C$ is the covariance matrix.
If the covariance of the data is independent of the model parameters $x_i$, as is the case for radio images, the above equation can be reduced to:
\begin{equation}
F_{i,j} = \frac{\partial{G}}{\partial x_i}C^{-1}\frac{\partial{G}}{\partial x_j} = \left(J^TC^{-1}J\right)_{i,j}
\label{eq:fim}
\end{equation}
where $J$ is the Jacobian.
Thus the new variance matrix is related to the original variance matrix but with a contribution from the (data) covariance matrix $C$.
Correct estimation of the error on each parameter means that we must replace the variance matrix returned by MINPACK, with the modified matrix $\left(J^TC^{-1}J\right)^{-1}$.
Aegean uses the covariance matrix to estimate the uncertainties according to eq\,\ref{eq:fim}.

We use the simulated test data (\S\,\ref{sec:data}) to compare the reported uncertainties with the deviation between the measured and true parameter values.
For a population of measurements the z-score, defined as: 
\begin{equation}
\frac{\Delta}{\sigma} \equiv \frac{\textrm{measured}-\textrm{true}}{\textrm{uncertainty}},
\end{equation}
will have mean of 0 and variance of 1 if the reported uncertainties are accurately and precisely estimated.
With the simulated data we have access to the true value of each parameter that is being fit for each source and thus it is possible to determine the z-score distribution
The z-score distribution can then be used to determine the accuracy to which the uncertainties are being reported.

\begin{figure}
\includegraphics[width=0.95\linewidth]{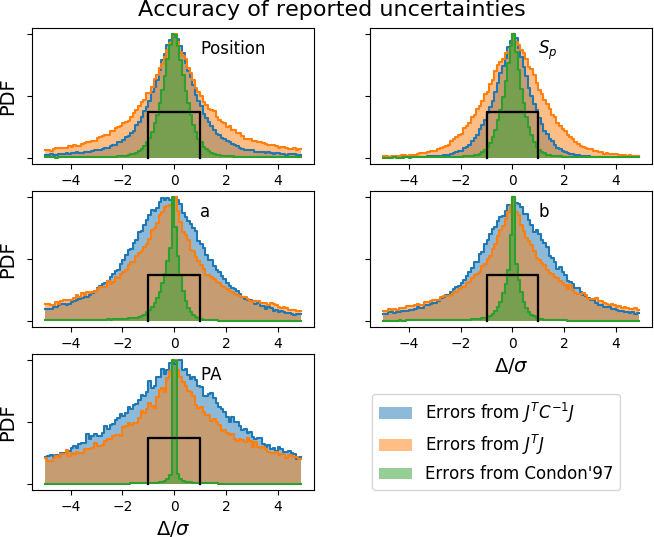}
\caption{
A comparison of the accuracy to which the uncertainties are reported by three different methods.
Blue/Orange distributions represent uncertainties derived from the FIM using Eq\,\ref{eq:fim} with or without the inverse covariance matrix.
The green distribution uses the method described by \citet{Condon1997}.
The black box indicates a standard deviation of units, which occurs when the uncertainties are accurately reported.
Distributions narrower than the black box indicate that the reported uncertainties are too large.
}
\label{fig:histograms}
\end{figure}

We explore three methods of calculating errors: using the Fisher information matrix (eq\,\ref{eq:fim}) with covariance matrix, using the Fisher information matrix without the covariance matrix (ie $C=I$), and the semi-analytic uncertainties derived by \citet{Condon1997}.
Aegean fits a model to the data in pixel coordinates, which is then transformed into world (sky) coordinates using the world coordinates system (WCS) module from AstroPy \citep{TheAstropyCollaboration2013}.
In the cases where the eq\,\ref{eq:fim} is used, the uncertainties are calculated in pixel coordinates, and then transformed into world (sky) coordinates.
\citet{Condon1997} describes the uncertainties in both coordinates, and here we use their equations 21 and 41 to calculate the uncertainties in the world coordinates directly, with a correction for the correlation between data points.
Figure\,\ref{fig:histograms} shows histograms of $\Delta/\sigma$ for the position, peak flux, and shape parameters for the three methods.

Figure\,\ref{fig:histograms} shows that the uncertainties in the shape parameters are not well reported in any of the three methods explored.
For the position angle, this can be partially explained by the fact that the position angle has a $2\pi$ ambiguity, and is also not defined for circular sources.
It is not yet understood why the uncertainties for the semi-major and semi-minor axes are not well described by any of the methods explored.
For the position and peak flux, the best method for estimating the uncertainties is to use the covariance matrix, whilst not using the covariance matrix will under estimate the uncertainty, and the description of \citet{Condon1997} will over estimate the uncertainty.

\subsubsection*{Parameter bias}
\cite{Refregier_noise_2012} derive the expected covariance and bias that occurs when using a least squares algorithm to fit data with a (non-linear) elliptical Gaussian model.
They report a parameter variance that is consistent with eq\,\ref{eq:fim}, and additionally report a parameter bias of:
\begin{equation}
b[x_i] = -\frac{1}{2}\left(F^{-1}\right)_{ij}\left(F^{-1}\right)_{kl}B_{jkl} + \mathcal{O}\left(SNR^{-4}\right)
\label{eq:bias}
\end{equation}
where $B_{ijk}$ is the bias tensor given by:
\begin{equation}
B_{ijk} = \sum_p \frac{\partial G}{\partial x_i}\frac{\partial^2G}{\partial x_j \partial x_k}
\label{eq:biastensor}
\end{equation}
where the subscript $p$ indicates summation over all pixels.
\cite{Refregier_noise_2012} show that even in the case of uncorrelated data, to second order in SNR: the best fit position parameters are covariant to a degree determined by the shape parameters, the position and position angle are not biased, the amplitude and major axis are biased high, and the minor axis is biased low.

In an earlier work \citet{Refregier1998} outline a calculation for the variance and bias of parameters of a non-linear model in the presence of correlated noise.
The expected bias is due to two factors: the correlated nature of the data, and the non-linear nature of the model being fitted to the data.
The inclusion of the inverse covariance matrix into the fitting process should remove the bias due to the correlated nature of the data, however the bias introduced by the non-linearity of the source model will remain.

\begin{figure}
\includegraphics[width=0.95\linewidth]{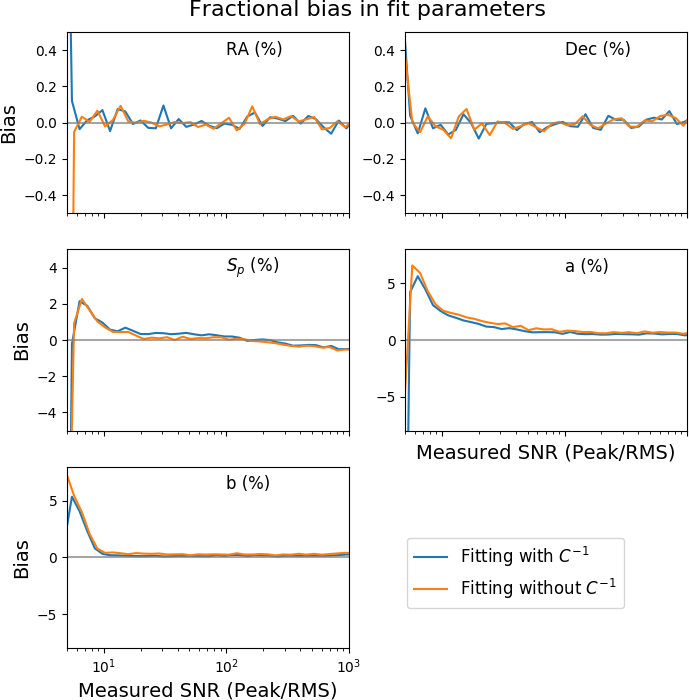}
\caption{
The bias in fitting each of the six parameters as a function of measured signal to noise ratio.
The peak flux density ($S_p$) has a small negative bias above about 1 Jy representing an underestimate of the true flux by about 1\%.
The major axis is biased high as low SNR and then low at higher SNR, whilst the minor axis is always biased high.
The RA, Dec, and position angle do not show any consistent biases.
The inclusion of the inverse covariance matrix reduces the bias for the major and minor axes at low SNR, but not by a significant amount.
}
\label{fig:bias}
\end{figure}

In Figure\,\ref{fig:bias} we measure the bias in five of the model parameters using the simulated data for both fitting methods.
The position angle bias is not shown. 
The bias is calculated as a fraction of the measured value and as a function of the measured SNR for the peak flux, semi-major and semi-minor axes. 
For the RA and Dec, the bias is reported as a fraction of the fitted semi-major axis.
We make the following observations:
\begin{itemize}
\item The position of the sources shows no bias in either RA or Dec to within 1\% of the semi-major axis.
\item The peak flux and the semi-major and semi-minor axes all show a positive bias of $2-5\%$ at $\textrm{SNR}<10$.
This bias is not related to the fitting process and is an example of the Eddington bias \citep{Eddington_fromula_1913}.
\item The peak flux density has a negative bias that is seen at a SNR of $10^2$, representing a fractional difference of $<0.1\%$.
\item The semi-minor axis shows no bias aside from the Eddington bias, whilst the semi-major axis shows a residual 1\% bias that persists to high SNR.
\item The 1\% bias in the semi-major axis will mean that the source area is also biased high, and that the calculated integrated flux will be biased.
\item The bias in the integrated flux (not shown) is also high at low SNR but asymptotes to 0 bias at high SNR, indicating that the peak flux and semi-major axis biases cancel out at high SNR.
\end{itemize}

Figure\,\ref{fig:bias} distinguishes between two fitting methods, with and without the inverse covariance matrix.
None of the parameters show significant difference in the bias when using or not using the inverse covariance matrix.
As noted previously, the bias due to the non-linearity of our source model cannot be recovered by using the inverse covariance matrix, whilst the bias due to correlated data can be.
Thus we conclude that the cause of bias in the non-linear least squares fitting is dominated by the effects of the non-linear source model.
This example demonstrates the asymptotic behavior of maximum likelihood estimators: The Cram\'er-Rao bound is met, and the estimator is unbiased, at high SNR but not necessarily at low SNR.
At low SNR the Eddington bias is dominant.
The ability to calculate and apply a correction for the bias induced by the spatial covariance of the data has been included in \aegean{} but is not enabled by default.

\section{Background And Noise Estimation - \bane{}}
\label{sec:bane}
Here we compare the background and noise estimation that is performed by \aegean, and that by \bane.
We denote the two algorithms as Zones (used by \aegean) and Grid (used by \bane).

The two algorithms described below share a number of design choices.
Firstly, the size of the zone in the Zones algorithm, and the box in the Grid algorithm, are chosen to have width and height that is 30 times the size of the synthesized beam. 
This choice has been shown by \citet{Huynh_completeness_2012} to optimize the completeness and reliability of the extracted compact source catalgoues. 
Secondly, the pixel distribution within a region is assumed to contain a contribution from a large scale background emission, a variance due to noise (a zero mean Gaussian distribution), and real sources (a roughly Poissonian distribution with a very long positive tail). 
The background and noise properties are typically calculated as the mean and standard deviation of the pixel distribution, however this neglects the contribution from astrophysical sources.
Each of these three components are assumed to vary across the image of interest.
The goal of the Zone and Grid algorithms is to estimate the slowly varying background component, the stochastic noise component, without knowledge of, or contamination from, the sources of interest.
In the presence of real sources, efforts need to be made to prevent the background and noise parameters from being biased.

\subsection{Zones algorithm}
\label{sec:bane:zones}
The background and noise estimation process that is performed by \aegean{} is based on a zones algorithm.
The zones algorithm divides an image into some number of zones and then computes the background and noise properties of each zone.
The pixels within a given zone are used to calculate the 25, 50, and 75th percentiles of the flux distribution.
The background is taken to be equal to the 50th percentile (the median), whilst the noise is taken to be equal to the inter-quartile range (IQR; 75th - 25th percentile) divided by 1.349 (corresponding to $1\sigma$ if the pixels follow a Gaussian distribution).
Calculating the RMS from the IQR range reduces the bias introduced by source pixels.
These background and noise properties are assumed to be constant over a zone, but can vary from zone to zone.
This approach is fast to compute, is simple to implement, but will not capture noise and background variations that vary on spatial scales smaller than the size of each zone.

\subsection{Grid algorithm}
\label{sec:bane:grid}
An alternative algorithm is implemented by \bane{} and it is similar to zones except that it takes a sliding box-car approach.
The grid algorithm works on two spatial scales: an inner (grid) scale, and an outer (box) scale.
The grid algorithm calculates the background and noise properties of all pixels within a box centreed on a given grid point.
The pixels within a given box are subject to sigma clipping, whereby the mean and standard deviation are calculated, values that are more than $3\sigma$ from the calculated mean are masked, and the process is repeated 2 times.
Such sigma-clipping reduces the bias introduced by source pixels, beyond that afforded by the IQR approach.
This is similar to the background calculation that is used by SExtractor \citep{Bertin_sextractor_1996}.
The next grid point is then selected and the process is repeated.
Since the grid points are separated by less than the size of the box, the process naturally provides a somewhat smoothed version of the Zones algorithm.
Once the background and noise properties have been calculated over a grid of points in the image, a linear interpolation is used to fill in the remainder of the pixel values. 
If the grid size is set to 1x1 pixels then this algorithm is equivalent to a box-car filter with sigma-clipping.
Since radio images are spatially correlated on scales of the synthesized beam, there is little loss of accuracy by increasing the grid size to be 4x4 pixels (a typical synthesized beam size.
This small loss of accuracy will then reduce the number of computations required by a factor of 16 - greatly increasing the speed of operation.
The grid algorithm is slower but more accurate than the Zones algorithm, the speed and accuracy can be balanced by adjusting the grid and box sizes.
In the case that the background is changing on spatial scales smaller than the box size, the background and noise properties cannot be calculated accurately in a single pass.
In such cases the background must first be calculated, and then the noise can be calculated from the background subtracted data.

The background and noise maps can be stored in a compressed format (not interpolated) and are automatically interpolated at load time by Aegean.
This compressed format saves a large amount of storage space, at a modest computation cost on load time.

\subsection{Algorithm comparison}
\label{sec:bane:compare}
\begin{figure}
\begin{center}
\includegraphics[width=0.95\linewidth]{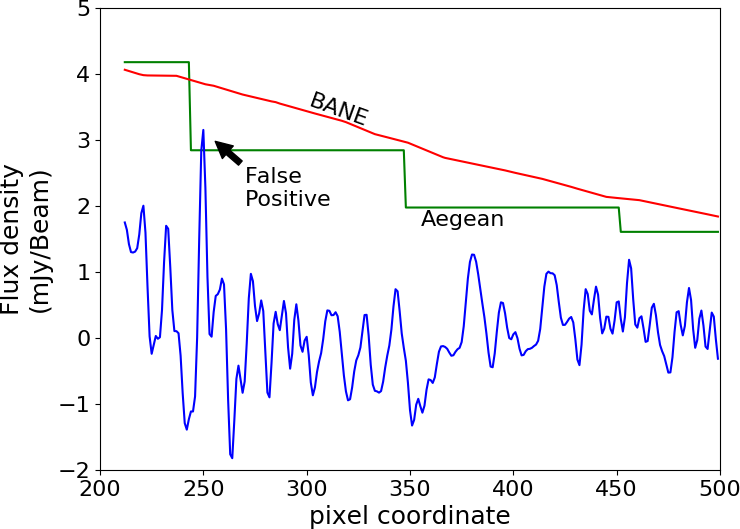}
\caption{
A demonstration of the difference between the \aegean{} and \bane{} background and noise maps on the resulting detection threshold.
The figure shows a cross section through an image along one of the pixel axes: flux density as a function of location within the image.
The blue line represents the image data.
The green and red lines represent the detection threshold (background + $5\sigma$) as calculated using \aegean{} and \bane{} characterizations of the background and noise. 
The difference in the two thresholding techniques results in a false positive when using the \aegean{} method, but no false positives when using the \bane{} method.
} \label{fig:zone_grid_ex1}
\end{center}
\end{figure}

The zones and grid algorithms are compared in Figure\,\ref{fig:zone_grid_ex1}, using observational data.
The example in question demonstrates that the zones algorithm is, at best, only accurate in the centre of each zone, and that towards the edge of the zone both the background and noise become incorrect.
The result of this error is to admit false detections at a rate that is in excess of what could be reasonably expected from simple Gaussian statistics.

\begin{figure}
\begin{center}
\includegraphics[width=0.95\linewidth]{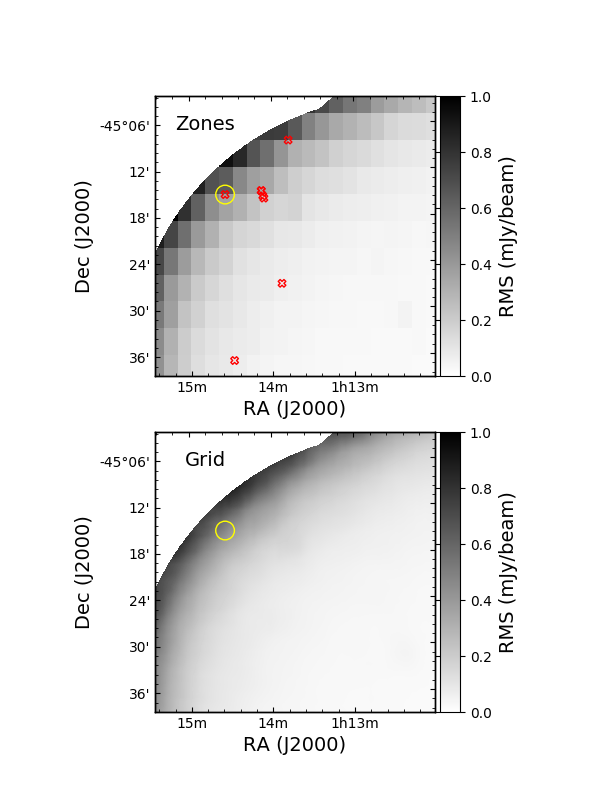}
\caption{
A comparison of two methods for calculating the RMS of an image. 
Upper: The noise map calculated using the zones algorithm, Lower: The noise map calculated using the grid algorithm. 
The red X's represent the location of spurious detections (false positives) due to inaccurate calculation of the background and noise characteristics of the image. 
The yellow circle denotes the false positive that is depicted in Figure\,\ref{fig:zone_grid_ex1}.}
\label{fig:rms_comp}
\end{center}
\end{figure}

In Figure\,\ref{fig:rms_comp} we demonstrate the false detections that are due simply to the inadequacies of the zones algorithm, using the observational test real data taken from a single epoch of the Phoenix deep field \citep{hopkins_phoenix_1998} studied by \citet{Hancock_radio_2016}.
Figure\,\ref{fig:rms_comp} shows that the false detections occur preferentially towards the edge of the image where the sensitivity is decreasing rapidly due to primary beam effects.
In this area the zones algorithm makes mistakes of the type depicted in Figure\,\ref{fig:zone_grid_ex1}, and close examination will show that the false detections are indeed at the boundaries of a zone.
The background and noise can also be misrepresented in interior image regions near to a bright sources or sources that are not able to be well {\sc clean}ed.
Since this local increase in noise is much smaller than the box over which the noise properties are calculated, there is an increased chance that side-lobes and clean artifacts will rise above the calculated local rms.
In a region approximately the size of the calculation box the local rms will be artificially high, meaning that even away from the troublesome source the completeness of the extracted catalogue will be reduced.
A typical approach to avoiding the problems of quickly increasing noise is to reduce the area of interest to exclude the outer regions of an image such as shown in Figure\,\ref{fig:rms_comp}.
Mitigating the effects of the false positives near bright sources in the interior of the image can be achieved with better $(u,v)$ coverage, more careful cleaning, or simply be excising a small area around the problematic sources.
These approaches are effective, however they reduce the sky area surveyed.
For a given amount of telescope observing time, a reduction in the sky area covered is equivalent to a reduction in sensitivity.
Better background and noise estimation is therefore equivalent to an increase in observing sensitivity.
For studies interested in detecting transient sources in the image domain, the reduction of false positives translates to a smaller number of false transient candidates, an increased confidence in the transients that are found, and reduced load on the research team.

\subsection{Caveats and future work}
In the GLEAM survey paper\citep{Hurley-Walker_galactic_2017}, it was noted that \bane{} was not correctly representing the noise properties of the images, and that this was due to the sigma-clipping that \bane{} implements.
Whilst this is true, it is not the whole story.
The GLEAM survey is sensitivity limited by a combination of side-lobe and classical confusion, resulting in a pixel distribution which is skewed towards positive values.
Even after sigma-clipping this skewed distribution means that the standard deviation that \bane{} calculates is not just the image noise, but a combination of the thermal noise plus a contribution from confusion.
The effects of confusion are reproduced in the simulated test image, where the number of sources per synthesized beam increase towards the south celestial pole.
In this region of the test image, \bane{} is not able to accurately reproduce the background and noise properties due to confusion.
Currently \bane{} makes little use of the WCS header information beyond determining the number of pixels per synthesized beam.
The grid/box size that is chosen by \bane{} is appropriate for the 'centre' of the image.
For sinusoidal (SIN) projected images this choice need not change as the beam sampling is typically constant over the entire image, however for large images and other projections this is no longer the case.
Indeed the simulated image that was described in section\,\ref{sec:data} has a point spread function that varies over the image, and so the beam sampling changes accordingly.
The result is that the grid/box size is not well chosen for the entire image, and there is a possibility that the spatial filtering will break down, and the separation of background, noise, and signal, can degrade.
This is an issue that is currently under development and will be addressed in future versions of \bane{}.


\section{Variable point spread function}
\label{sec:psf}
A typical radio image has a point spread function ({\sc psf}) which is equal to the synthesized beam, and which is constant across the field of view.
At frequencies below $\sim 150$\,MHz, the ionosphere can induce a lensing effect which can decouple the {\sc psf} from the synthesized beam in a manner similar to seeing in optical images \citep[as seen by][]{Loi_new_2016}.
Additionally, stacking or mosaicking of images which are taken under different ionospheric conditions can introduce a blurring effect, due to uncorrected ionospheric shifts \citep[as seen by][]{Hurley-Walker_galactic_2017}.
A radio interferometer will have a synthesized beam that changes with elevation angle.
In a SIN projection with the observing phase centre at the projection centre, the sky to pixel mapping and the synthesized beam will both transform in the same way, at the same rate, and thus the synthesized beam will remain constant in pixel coordinates.
For small fields of view the synthesized beam can be approximated as constant.
However for large fields of view, one or more of the above effects will result in a position dependent point spread function that must be accounted for.
Failure to account for a direction dependent point spread function will result in a biased integrated or peak flux measurement, depending on how the flux calibration is calculated.
In order to achieve a proper accounting of the peak flux and shape (and thus integrated flux) of a source over the full filed of view, source characterization must be able to use a variable point spread function.

\begin{figure}
\centering
\includegraphics[width=0.95\linewidth]{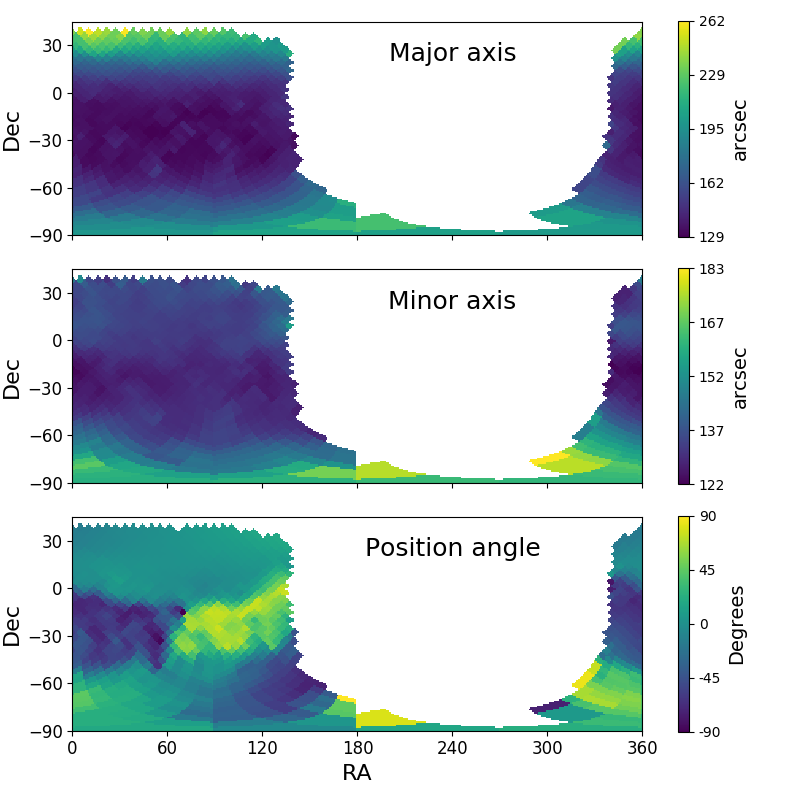}
\caption{
An example {\sc psf} map demonstrating the variation of semi-major axis size as a function of position on the sky.
The observations contributing to this image are meridian drift scans and thus the semi-minor axis of the synthesized beam should not change with zenith angle.
The variations that are seen here are due to differing ionospheric conditions and a blurring effect that is introduced in the mosaicking process.
These data drawn from \citep{Hurley-Walker_galactic_2017}
}
\label{fig:psf}
\end{figure}

In order to allow \aegean{} to incorporate a variable point spread function we have implemented a {\sc psf} model  that works in one of two ways.
For images that don't suffer from ionospheric blurring the shape of the synthesized beam can be calculated from the zenith angle, which in turn can be calculated from the latitude of the array.
In this case a user only needs to indicate either the telescope being used or its latitude.
\aegean{} `knows' the latitude of many radio interferometers and so the user can for example use the option \verb+--telescope=MWA+.
For images that have a position dependent {\sc psf} that is not simply zenith angle dependent, \aegean{} can be supplied with an auxiliary map that gives the semi-major/semi-minor axes and position angles over the sky.
The {\sc psf} map is a FITS format image, with 3 dimensions.
The first two dimensions are position (RA/Dec), and the final dimension represents the shape of the {\sc psf} in degrees (semi-major, semi-minor, PA).
There is no need for the {\sc psf} map to be in the same projection or resolution as the input image.
Figure\,\ref{fig:psf} shows the {\sc psf} map for one week of the GLEAM survey.
The AegeanTools package currently doesn't provide any mechanism by which a {\sc psf} map can be produced, however see \citep{Hurley-Walker_galactic_2017} for an example of how this can be done.

To the best of our knowledge \aegean{} is currently the only source finding software that is capable of implementing this feature, despite the fact that it will be critically important for source characterization with the SKA-Low.


\section{Priorized fitting}
\label{sec:priorized}

Fitting uncertainties are significantly reduced when the number of free parameters are also reduced.
If a source is known to exist at a given location, then a user may want to ask: ``What flux is consistent with a source with a given location and shape?"
The traditional approach of recording an upper limit makes statistical analysis difficult, and does not use all of the available data.
A new method has been implemented that will allow source characterization to be achieved independent of the source finding stage.
This process is analogous to aperture photometry in optical images, and since it relies on prior information, we use the term {\em priorized} fitting.
Priorized fitting will result in measurements with associated uncertainties, rather than a mix of measurements and limits, making it possible to use a greater variety of statistical methods when analyzing the data.

\begin{figure}
\centering
\includegraphics[width=0.9\linewidth]{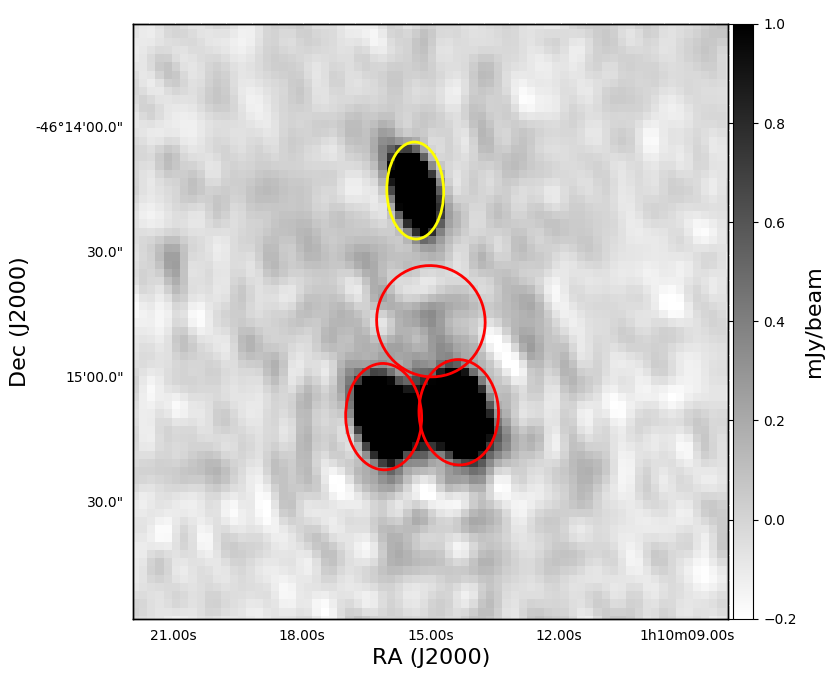}
\caption{
An example of the source regrouping that is performed by \aegean{} to ensure that overlapping sources are jointly fit.
An ellipse represents the location and shape of each component. 
Three components in the red/lower island are jointly fit in both the blind and priorized fitting method. 
The yellow/upper component is fit separately.
}
\label{fig:regions}
\end{figure}

When two or more sources are near to each other they can become blended.
When fitting for the flux of a source that is near to another source, the fitted flux will be biased.
In order to avoid this blending bias, sources which are near enough to become blended are grouped and jointly fit.
By default, sources which overlap each other at the half power point (have overlapping ellipses) are put into the same group.
The model that is fit contains all the sources within a group.
Figure\,\ref{fig:regions} shows an example of the regrouping and priorized fitting.

\begin{figure}
\centering
\includegraphics[width=0.9\linewidth]{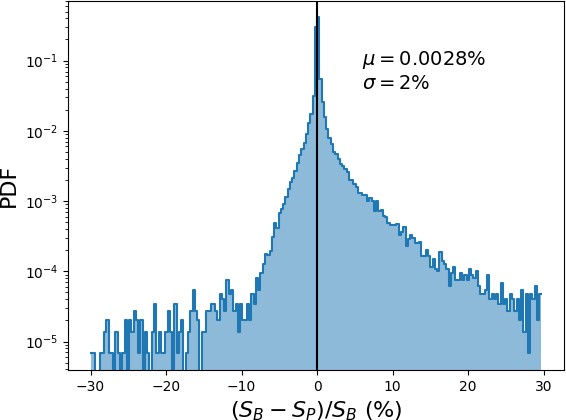}
\caption{
A comparison of the peak flux as measured by \aegean{} in the blind source finding mode $S_B$ or the priorized fitting mode $S_P$, using the simulated test image.
The fluxes agree to within their respective uncertainties, with a variance of just 2\%.
}
\label{fig:priorized}
\end{figure}

The pixels that are included in the fit are selected based on their distance from the source to be fit and the size of said source.
The selection of pixels for priorized fitting is thus different from that which occurs during the normal blind find/characterize operation of Aegean.
This choice of pixels allows for sources that are below a nominal $5\sigma$ detection threshold to be measured.
For this reason it is possible for the priorized fitting routine to return a negative flux value.
Sources that are poorly fit initially, or not well described by an elliptical Gaussian will be poorly measured by a priorized fit.
Due to the possibility that different pixels may be used in the blind and priorized fitting, the resulting fluxes may differ.
However, our tests show that sources that are initially well fit have a priorized peak flux that is identical to within the reported errors.
Figure\,\ref{fig:priorized} compares the flux that is measured by \aegean{} in a blind source finding mode (the default), as compared to that measured with the priorized fitting mode where only the flux is fit.

Priorized fits are treated in the same way as the regular (blind) detections.
This means that by default the inverse covariance matrix is used in the fitting, the errors are measured according to Sec\,\ref{sec:nlls:errors}, and the results are reported in the same tabular formats.
\aegean{} generates a universally unique identifier (UUID) for each source in the blind source finding stage, and then copies this UUID during the priorized fitting.
Thus the light curves or spectral energy distributions can be easily reconstructed by doing an exact match on the UUID key.
Matching on UUID instead of position will avoid the many problems and uncertainties associated with cross-matching catalogues.

Since priorized fitting is a two step process (find and then remeasure) it is now possible for \aegean{} to use one image as a detection image, and use a separate image as a measurement image.
The following use cases come immediately to mind: variability, spectral studies, and polarization work.
In searching for radio variability a deep image can be created with all the available data from which a master catalogue can be created.
Producing light curves for all the sources present in the deep image is then simply a matter of doing a priorized fit on each image from each epoch, using the master catalogue as an input.
A similar approach can be taken when working with images at multiple frequencies: a single image is chosen as the reference image, and then source fluxes are measured in each of the other images, producing a continuous spectral energy distribution for every source.
This is the approach taken for the GLEAM survey \citep{Hurley-Walker_galactic_2017}.
Finally, even for a single epoch and frequency, priorized fitting can be used to measure polarized intensity in stokes Q, U and V images, for sources detected in a stokes I image.

In each of the use cases outlined above, the advantage of priorized fitting is that every source of interest is assigned a measurement and uncertainty.
This means that the resulting light curves, spectral energy distributions, or polarization states, do not contain limits or censored data.
The measurement of source flux may become negative in the low SNR regime, and while this may not be physically meaningful, it is statistically meaningful, and such measurements should be included, for example, when fitting a model spectral energy distribution \citep{Callingham_extragalactic_2017}.

\citet{Chhetri_interplanetary_2017} recently used \aegean{} to find sources in an image representing the standard deviation of a data cube.
These variable sources were then characterised by using priorized fitting to extract their mean fluxes from an image formed from the mean of the data cube.
In this way, the modulation index of the scintillating component was able to be separated from non-scintillating components in sources which may have arc minute scale structure.


\section{Other source models}
\label{sec:island}
\aegean{} was designed to find and characterize compact sources.
\aegean{} identifies islands of pixels using a signal to noise threshold.
This threshold is applied on the absolute value of the SNR, and thus both positive and negative sources are identified in the source {\em finding} phase.
By default only sources with positive SNR are characterized and reported, but \aegean{} is also able to characterize negative sources.
The option \verb+--negative+ will turn on this feature and allow, for example, both left and right circularly polarized sources to be identified in Stokes V images, in a single pass.

\begin{figure}
\centering
\includegraphics[width=0.9\linewidth]{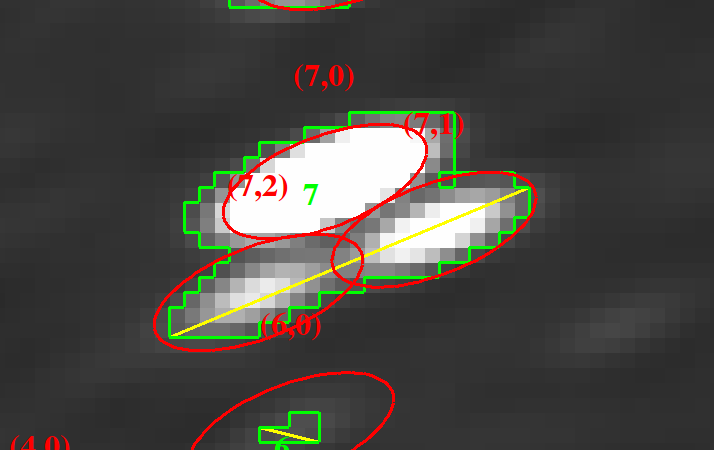}
\caption{
Example of the island characterization that is available in \aegean{}.
DS9 is used to visualize the extent and location of the island.
The red ellipses show the components that were fit with (island, component) labels.
The green borders show the pixels that were included in each island with label of the island number.
The island number in the islands catalogue can be used to identify which components were fit to this island from the components catalogue.
The yellow line indicates the largest angular extent of the island.
}
\label{fig:island}
\end{figure}

\aegean{} finds sources based on islands of pixels \citep{Hancock_compact_2012}, fitting one or more components to each of these islands.
Diffuse or resolved sources are not well fit by \aegean, however the islands that are identified can be characterized in terms of their position, area, angular extent, and integrated flux.
The option \verb+--island+ will cause \aegean{} to also report the parameters of pixel islands, both positive and negative.
Island properties are written to a separate catalogue, which can be linked to the source catalogue using the {\em island} column.
For both island and component catalogues, \aegean{} can write a DS9 region file that identifies exactly which pixels within the image contributed to each island.
Figure\,\ref{fig:island} shows an example of a DS9 visualization of an island that was characterized by \aegean.

\section{Sub-image searching with \mimas{}}
\label{sec:regions}
\begin{figure}
\centering
\includegraphics[width=0.9\linewidth]{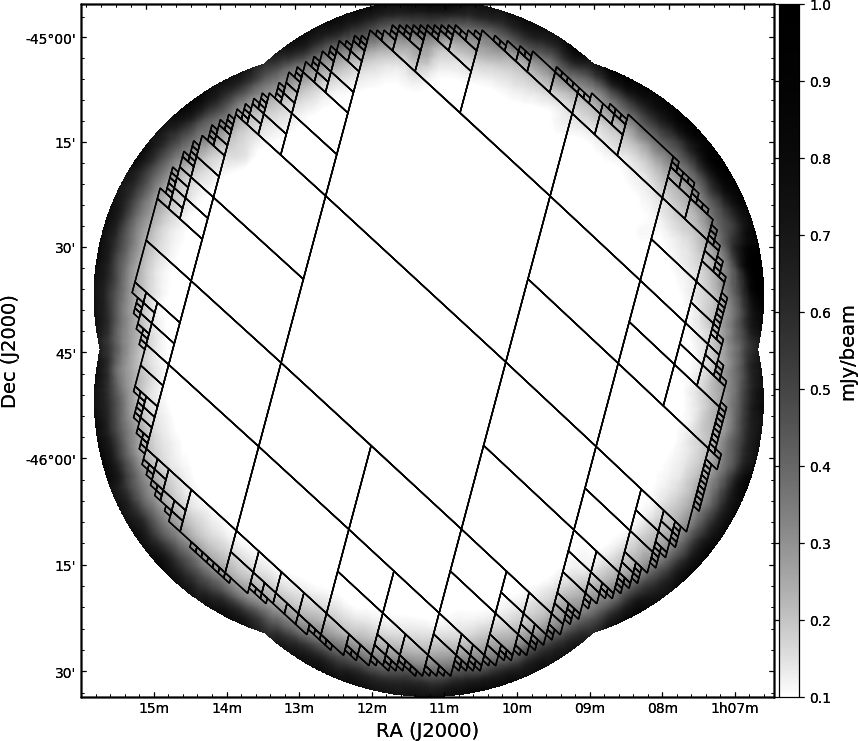}
\caption{
Example use of a region mask to constrain the area over which source finding will be performed.
The background image is an rms map generated by \bane{}, with a linear color scale from $0.1-1$\,mJy/beam.
The black diamonds show the masking region, which are represented by HEALPix pixels of different order.
Only islands of sources which overlap the mask region are fit by \aegean{}.
}
\label{fig:mimas}
\end{figure}

\aegean{} suffers from a defect that occurs when a large region of an image is included in a single island.
The covariance matrix grows as the square of the number of pixels within an island, and the number of sources also increase.
Both of these effects cause the fitting of an island to take a very large amount of time, and can cause a crash.
There are two solutions to this problem.
The first solution is to mask the pixels in an image which would cause a large island to be found.
This is typically in regions towards the edge of an image where the noise becomes high, near bright sources that are not able to be {\sc clean}ed completely, or around extended or resolved sources such as within the Galactic plane.
A second solution is to leave the image untouched, but to provide a masking file to \aegean{}.
The masking file will cause \aegean{} to ignore any islands whose pixels don't fall within the masking region.
The second method has the advantage that these regions can be calculated in advance, and a single such mask can be used for many images.
By separating the image from the mask, it also means that users no longer need to have masked and un-masked versions of their data on disk.
This is the method used by \citet{Meyers_southern-sky_2017} to avoid finding sources near the edge of their survey images.
The format of the mask file is a pickle of a python object that is created using the Multi-resolution Image Masking tool for Aegean Software (\mimas).
Figure\,\ref{fig:mimas} shows an example of the use of such a mask region to exclude the high noise regions from the observational test image.

\mimas{} uses a Hierarchical Equal Area isoLatitude Pixelization of a sphere \citep[HEALPix,\footnote{\href{http://healpix.sourceforge.net/}{healpix.sourceforge.net}}][]{Gorski_healpix_2004}, to represent the sky as a set of pixels.
Storing sky areas as sets of HEALPix pixels make is possible to combine regions using binary set operations.
Currently \mimas{} supports union and difference operations from the command line, but the underlying module ({\sc AegeanTools.Regions}) is able to support all the set operations provided by the built in {\sc set} class.
Python pickle files are not amenable to easy visualization, so \mimas{} provides two methods for visualization.
First is a Multi-Order Coverage map (MOC\footnote{\href{http://www.ivoa.net/documents/MOC/}{www.ivoa.net/documents/MOC}}) as a .FITS file.
These MOC files can be easily visualized and manipulated by the Aladin viewer \citep{Bonnarel2000}.
A second method of visualization is a region file that is readable by DS9.


\section{Summary}
\label{sec:summary}
We have addressed the issue of fitting non-linear models to correlated data as it applies to radio astronomy images.
We have developed a method that accounts for the correlated nature of the data in the fitting process, but found that the resulting fit parameters were not significantly different as a result.
The reported parameter uncertainties and calculated biases were presented.
We find that including the data covariance matrix in the Fisher information matrix gives the best estimate of the uncertainty in position and peak flux, whilst none of the three methods investigated were able to accurately report uncertainties for the shape parameters.
The parameter biases that we detect are dominated by the non-linear nature of the source model and not the data covariance.
\aegean{} has been modified to use the data covariance matrix in the reporting of parameter uncertainties.

We presented an algorithm for estimating the background and noise properties of an image and compared it with more simplistic methods currently in use.
We find that the background and noise images created by \bane{} result in a lower false detection rate, especially in the case where the background or noise properties are changing quickly within an image.

We presented a method of priorized fitting that allows for a more statistically robust estimate of the flux of a source even when the source is below the classical detection threshold.
This priorized fitting simplifies the analysis of light curves and spectral energy distributions by replacing upper limits with a statistically meaningful measurement and uncertainty.
\aegean{} is able to perform priorized fitting with a choice of the number of degrees of freedom, and includes a regrouping algorithm that ensures that overlapping sources and components are fit jointly.

Wide-field imaging requires that the local point-spread function be allowed to vary across the image in a possibly arbitrary manner.
We have provided description of how to create a FITS format image that will describe the changing PSF, and \aegean{} is able to use such an image to correctly characterize sources.

We have developed a method for describing regions of sky of arbitrary complexity, based on the HEALPix projection of the sphere.
This method is made available via the \mimas{} program, and the region files that it can produce can be used to constrain the area of sky over which \aegean{} will find sources.

The overall development path for \aegean{} and \bane{} has been driven by the current needs of radio astronomers and the anticipated future needs of astronomers working on the SKA.

\section{Future development plans}
\label{sec:future}

In order to make better use of the multiple cores available on desktop and HPC machines, \aegean{} has been modified to spread the process of fitting across multiple cores.
\bane{} was created with a parallel-processing capability from the outset.
The multi-processing for both \aegean{} and \bane{} is currently made possible via the {\sc pprocess} module\footnote{\href{https://pypi.python.org/pypi/pprocess}{pypi.python.org/pypi/pprocess}}. 
Spreading the processing across multiple cores is done by forking, and thus the memory usage is multiplied by the number of processes, and there is no capability for spreading across multiple computing nodes within an HPC environment.
Work is underway to migrate to an OpenMPI\footnote{\href{http://www.open-mpi.org/}{www.open-mpi.org}} based approach which will reduce the total memory usage, and allow the processing to be spread across multiple nodes within an HPC environment.
With the many new HPC facilities offering GPU nodes as well as CPU nodes there is significant motivation for a GPU implementation of both \aegean{} and \bane{} and expertise is being sought for such an implementation.

\bane{} currently works with a square grid that is constant over an image.
The ideal grid size is dependent on the image PSF \citep{Huynh_completeness_2012} and so we are working on a method by which the grid and box size that is used by \bane{} will be able to also scale with the image PSF.
This development will further improve the performance of \aegean{} via more accurate background and noise models.

The intended use of \aegean{} is for continuum images and thus works only on a single image at a time.
Source finding and source characterization are two distinct tasks, and can be performed on separate images.
We plan to develop such a capability for \aegean{} such that source finding can be completed on a detection image, and then characterization on a separate image, or sequence of images. 
This will be a hybrid of the current blind finding/characterization and priorized fitting that \aegean{} is able to achieve.

The current ideology that is adopted by \aegean{} and \bane{} is that the background, noise, and sources are all independent of each other.
This is true of compact continuum images which have been well cleaned.
However image of polarized emission are inherently positive definite, and have a non-Gaussian noise distribution, whereby the noise and signal are not combined linearly, but in quadrature.
Thus the true estimation of the image noise requires knowledge of the sources within the image.
This suggests that the background and noise estimation needs to be performed before or in conjunction with the source finding and characterization process.

A common user request is for \aegean{} to be able to find sources in image cubes similar to the capability of Duchamp \citep{Whiting2012}.
An adjustment to the source model to include a spectral index, and possibly spectral curvature is a first step towards meeting this goal.
Image cubes that have a PSF that changes significantly with frequency are now being produced by instruments such as the MWA, which have a large fractional bandwidth.
A PSF that changes with frequency can be characterized in a manner similar to that described in section \ref{sec:psf}, by adding an additional dimension to the data.

\begin{acknowledgements}
We acknowledge the work and support of the developers of the following following python packages: Astropy \citet{TheAstropyCollaboration2013}, Numpy \citep{vaderwalt_numpy_2011}, Scipy \citep{Jones_scipy_2001}, LmFit \citep{newville_lmfit_2017}.

Development of \aegean{} made extensive use of the following visualisation and analysis packages: DS9\footnote{\href{ds9.si.edu}{http://ds9.si.edu/site/Home.html}}, Topcat \citep{Taylor_topcat_2005}, and Aladin \citep{Bonnarel2000,Boch_aladin_2014}.

We thank the following people for their contributions to \aegean{} whether they be in the form of code, bug reports, or suggested features:
Ron Ekers, 
Jessica Norris,
Sharkie,
Tara Murphy,
Shaoguang Guo,
Stefan Duchesne,
rmalexan,
Josh Marvil,
David Kaplan,
Yang Lu,
Robin Cook,
Martin Bell,
John Morgan,
Christopher Herron,
Sarah White,
Rajan Chhetri


This research made use of Astropy, a community-developed core Python package for Astronomy \citep{TheAstropyCollaboration2013}.
This research has made use of the VizieR catalogue access tool, CDS, Strasbourg, France. The original description of the VizieR service was published in \citet{vizier}.
This research has made use of ``Aladin sky atlas" developed at CDS, Strasbourg Observatory, France.
Parts of this research were conducted by the Australian Research Council Centre of Excellence for All-sky Astrophysics (CAASTRO), through project number CE110001020.
This work was supported by resources provided by The Pawsey Supercomputing Centre with funding from the Australian Government and the Government of Western Australia
This project was undertaken with the assistance of resources and services from the National Computational Infrastructure (NCI), which is supported by the Australian Government

\end{acknowledgements}

\nocite*{}
\bibliographystyle{pasa-mnras}
\bibliography{aegean2}

\appendix{}
\section*{Additional Software provided by AegeanTools}
\aegean{}, \bane{}, and \mimas{}, are all part of the AegeanTools library.
There are additional scripts available as part of this library that are useful and are discussed briefly below.

\section{\aeres}
\aeres{} is a program that will compute the a residual map when given an input image and a catalogue of sources.
\aeres{} was created to help test and verify the performance of \aegean{} but has been found to be useful for other purposes, and has thus been made available as part of the AegeanTools package.
The intention is that the input catalogue was created by \aegean{} on the input image.
In order to reduce the computational cost of modeling sources source models are only computed over a small sub-set of the entire image.
The sources can be modeled down to either a given fraction of their peak flux, or to a given SNR (default is SNR=4), a choice which can be controlled by the user with the \verb+--frac+ or \verb+--sigma+ options.

Alternative uses for AeRes include the ability to insert model sources into an image using the \verb+--add+ option.
The simulated test image discussed in section \ref{sec:data}, was constructed in this manner.
Alternatively, sources can be masked (pixels set to blank) from an input image using the \verb+--mask+ option.

Not all of the columns from the \aegean{} catalogue format are used by \aeres{}.
For users wishing to create their own catalogue outside of \aegean{}, the following columns are required:
\begin{itemize}
\item ra (degrees)
\item dec (degrees)
\item local\_rms (Jy)
\item peak\_flux (Jy)
\item a (arcsec)
\item b (arcsec)
\item PA (degrees)
\end{itemize}
All other columns may be ignored or set to Null values.

\section{SR6}
As mentioned in section \ref{sec:bane}, BANE is able to output compressed versions of the background and noise maps.
These maps are significantly smaller than the normal output maps, and differ only in the fact that the final interpolation has not been performed.
SR6 (Shrink Ray 6), is a helper tool that was initially created to enable a user to take a background or noise map created by BANE and convert between the compressed and non-compressed versions.
The decompression of an already compressed file is done using linear interpolation between pixels on a grid.
The compression of a map is implemented as decimation, where by every Nth pixel in a grid is saved.
The parameters of the initial image and the the compression state are stored in the FITS header with custom keywords of the form \verb+BN_XXXX+.
These same keywords are used when \bane{} is instructed to write a compressed output.
\aegean{} is able to recognize these keywords upon loading a file and, when present, the interpolation of a background or noise image will be done at load time.

\end{document}